\documentclass{article}
\usepackage{spconf,amsmath,graphicx}
\usepackage[export]{adjustbox}
\usepackage{subfigure}
\usepackage{url}
\usepackage{booktabs}
\usepackage{bm}
\usepackage{caption}

\title{Fine-grained Emotion Strength Transfer, Control and Prediction for Emotional Speech Synthesis}
 \name{Yi Lei, Shan Yang, Lei Xie\sthanks{Corresponding author. This work was supported by the National Key Research and Development Program of China (No.2017YFB1002102).}}
 \address{Audio, Speech and Language Processing Group (ASLP@NPU), School of Computer Science, \\
          Northwestern Polytechnical University, Xian, China}

\begin{document}
%
\maketitle
\begin{abstract}



This paper proposes a unified model to conduct emotion transfer, control and prediction for sequence-to-sequence based fine-grained emotional speech synthesis. Conventional emotional speech synthesis often needs manual labels or reference audio to determine the emotional expressions of synthesized speech.  Such coarse labels cannot control the details of speech emotion, often resulting in an averaged emotion expression delivery, and it is also hard to choose suitable reference audio during inference. To conduct fine-grained emotion expression generation, we introduce phoneme-level emotion strength representations through a learned ranking function to describe the local emotion details, and the sentence-level emotion category is adopted to render the global emotions of synthesized speech. With the global render and local descriptors of emotions, we can obtain fine-grained emotion expressions from reference audio via its emotion descriptors (for transfer) or directly from phoneme-level manual labels (for control). As for the emotional speech synthesis with arbitrary text inputs, the proposed model can also predict phoneme-level emotion expressions from texts, which does not require any reference audio or manual label.

\end{abstract}

\section{Introduction}
\label{sec:intro}
Thanks to the rapid development of deep learning, speech synthesis has been significantly advanced~\cite{ze2013statistical,qian2014training,ling2013modeling}. Recently, with unified acoustic and duration modeling, sequence-to-sequence (seq2seq) based neural speech synthesis has achieved superior performance with extraordinary naturalness compared to the conventional methods~\cite{wang2017tacotron,shen2018natural,li2019neural,yang2020localness}. Since natural-sounding can be reasonably produced by current seq2seq-based speech synthesis models learned from a typical corpus with neutral speaking style, there have been increasing interests in how to deliver expressive speeches with these seq2seq models~\cite{skerry2018towards,wang2018style,stanton2018predicting,zhang2019learning}. As we know, human speech is expressive in nature, with rich style expressions and subtle emotions~\cite{wang2018style}.

To achieve expressive speech synthesis, a common solution is to learn style-related latent representations from reference audio~\cite{skerry2018towards,wang2018style,bian2019multi,henter2018deep}. The goal is to make the synthesized speech to imitate the style of the reference audio, which can be treated as some kind of style transfer. Although it is possible to control the speaking style by analyzing and controlling the learned style representations~\cite{um2020emotional,wang2018style}, it is still hard to choose a suitable control vector for arbitrary sentences. Besides, these methods usually encode the reference audio into a fixed-length style representation without explicitly considering the length and contents of the target synthesized speech.  When aggregating such reference speech into a fixed-length embedding, essential temporal information may be lost. Thus they can only obtain a global or averaged style~\cite{bian2019multi}. Undoubtedly, human speech contains subtle expressions at various granularities. For instance, phoneme-level prosody variations are important for expressive speech synthesis~\cite{lee2019robust}.

This paper proposes a fine-grained control and prediction approach for emotional speech synthesis -- a typical case of expressive speech synthesis specifically focusing on emotion rendering. The emotional expressions in human speech are directly affected by their intentions, which leads to different emotion categories such as happy, angry and fear, or even different emotion strengths in each word or phoneme. For example, an angry speaker may particularly put a strengthened focus with strong intensity on `hate' when speaking ``I hate this!". Therefore, in this paper, we treat the emotion category as a \textit{global render} of speech, while the emotion strength in each word or phoneme is defined as a \textit{local descriptor}.  Modeling the global render is straight-forward: we directly adopt emotion embeddings to control the emotion category. So for the fine-grained emotional speech synthesis, the key problem is how to model and control the local emotion descriptors.

Since the explicit annotations of emotion strength are unavailable, we adopt a relative ranking method~\cite{ferrari2008learning,parikh2011relative} to represent the local emotion descriptors. In detail, we automatically learn the \textit{relative attributes} of emotional speech compared to the neutral speech in the utterance level, where the attributes are proved to be strength-related~\cite{zhu2019controlling}. In order to achieve fine-grained emotion control, we then extract the emotional strength in the phoneme level from the learned ranking function. The phoneme-level local descriptors are then utilized along with text inputs to build the seq2seq acoustic model. Experiments show that it is easy to control the global emotion render as well as the local emotion descriptor of each phoneme from either speech or manual labels.

Besides, even though we could easily control the global render and local descriptor through the above approach, it still needs a method to automatically choose a suitable control vector for arbitrary text inputs during inference~\cite{stanton2018predicting}. To build an easy-to-use fine-grained emotional speech synthesis model, we further introduce a prediction module in the acoustic model to predict the local descriptors. With such a prediction module, the acoustic model has the ability to produce natural emotional speech according to the contents of the input text. As another advantage, our model can also accept control vectors from reference speech or manual labels to conduct style transfer and control at the same time. The efficacy of the proposed approach is validated from experiments.

\section{Related Works}
\label{sec:format}

Recently there are various attempts to model emotions or styles in speech synthesis~\cite{skerry2018towards,wang2018style,bian2019multi,um2020emotional,lee2019robust,lorenzo2018investigating,lee2017emotional}. The most straightforward way to conduct emotional speech synthesis is using explicit annotations or labels as global render to model expressions~\cite{lorenzo2018investigating,lee2017emotional}.  But the global explicit constraints can only provide an ``averaged'' emotional voice~\cite{stanton2018predicting}, and it is also hard to flexibly control the local emotion variations or even global intensities. To continuously control the global emotion render, the method in~\cite{zhu2019controlling} introduced a ranking function to represent the emotion strength of speech. In this way, emotional speech can be produced with different strengths. However, it can only control the global render, i.e., global strength, and manual instructions are also necessary to decide the strength of synthesized speech during inference.

To avoid explicit labels, the approach in~\cite{skerry2018towards} adopts a reference encoder to extract a latent prosodic representation for style imitation. But the performance of synthesized speech is directly affected by the choice of specific reference audio during inference. Based on the reference encoder method, Global Style Token (GST) is proposed to learn interpretable style embeddings in an unsupervised way~\cite{wang2018style}.  With GST, the proposed model could imitate the style of reference audio and control the style of synthesized speech by choosing specific tokens.  As for its application in emotional synthesis, the approach in~\cite{um2020emotional} introduces an inter-to-intra distance ratio algorithm on the learned style tokens, which minimizes the intra-cluster embedding vectors and maximizes the inter-cluster ones. An interpolation technique is further proposed to control emotion intensity. But the above methods still require reference audio or manual labels to guide the generation process and have not explicitly considered controlling local subtle emotion expressions.

Without the need for auxiliary feature during inference, TP-GST is proposed to predict expressive speaking style directly from text~\cite{stanton2018predicting}. It uses a pre-trained GST model to extract global style tokens of each utterance in the training data, and there is a second task in the encoder to predict style tokens from texts. Again, the TP-GST method only models the global style render of synthesized speech and it is still hard to conduct fine-grained emotional control. In this paper, we focus on fine-grained emotional speech synthesis, which considers the global render and local emotion descriptors at the same time. After extracting granularized phoneme-level emotion strength through the ranking function, we model and predict the local emotion descriptors in a unified model. In this way, the model can generate fine-grained emotional speech directly from text without manual control, and we can also control the generation process through granularized constraints from manual labels or reference audio during inference.

\section{Proposed Model}

\begin{figure*}[ht]
        \centering
        \includegraphics[width=1.0\linewidth]{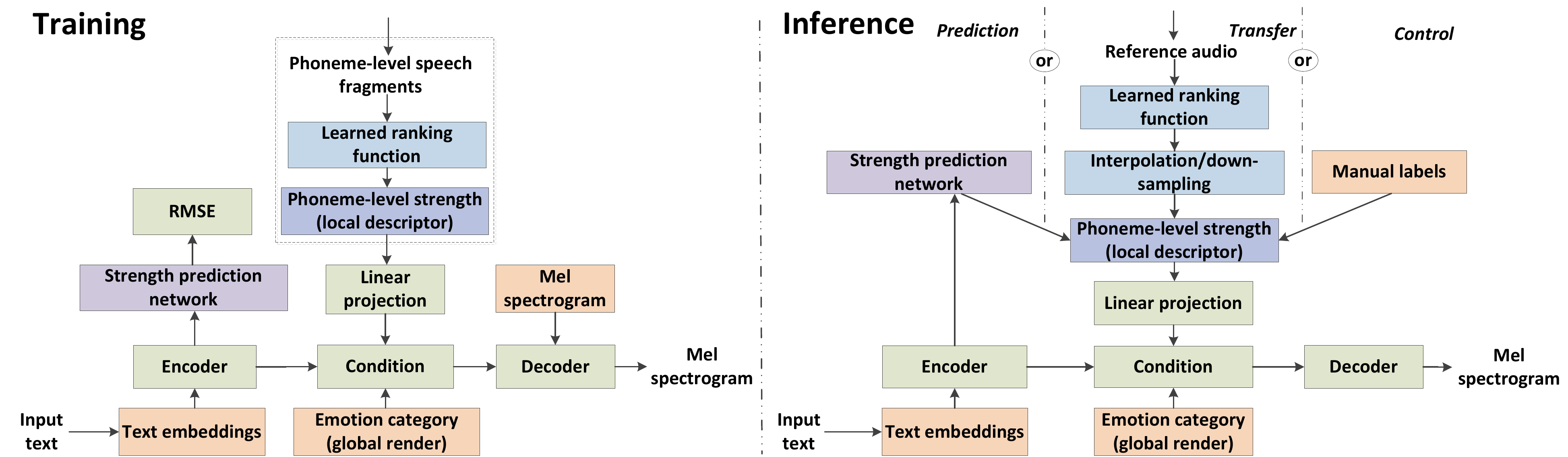}
        \caption{System overview}
         \vspace{-10pt}
        \label{fig:sys}
\end{figure*}

Fig.~\ref{fig:sys} shows the proposed fine-grained emotional speech synthesis framework for transfer, control and prediction.  It shares similar architecture with Tacotron~\cite{wang2017tacotron} and Tacotron2~\cite{shen2018natural}, which is composed of a CBHG-based text encoder and an attention-based auto-regressive acoustic decoder to generate mel-spectrogram. As for the emotion expression modeling, the proposed model contains a flexible module to provide emotional information during speech generation, which is learned from text inputs (prediction) or extracted from reference audio (transfer) or manual labels (control). With the emotion expression module, we can conduct emotion transfer, control and prediction in a unified model.

\subsection{Local emotion descriptors extraction}
For flexible fine-grained emotion speech synthesis, the key aspect is how to represent the local emotion expressions. In our work, we aim to learn a phoneme-level emotion strength representation to conduct fine-grained emotion modeling. Since the phoneme-level emotion descriptors are not readily available and it is hard to annotate manually, we use the concept of relative attributes~\cite{parikh2011relative,zhu2019controlling} to learn the emotion strength of phonemes to achieve this goal.

Given two categories of data, the ranking function is to calculate the relative attribute of the data~\cite{parikh2011relative}. In this paper, we treat the emotion strength as an attribute of speech. Hence the ranking function aims to learn the relative difference of emotion strength between neutral speech and a kind of emotional speech (such as happy). With the learned ranking function, we can obtain the relative strength for unseen emotional speech, which we treat as local emotion descriptors in this paper.

Assuming the training set for learning the ranking function is $T$ = \{\emph{t}\} represented in \(R^{n}\) by acoustic features \{\(x_{t}\)\}, and $T$ = $N$ \(\cup\) $H$, where $N$ and $H$ are the neutral and happy emotion set, respectively. The goal of relative attributes is to learn ranking function
\begin{equation}\label{eq:1}
 r( \bm{ x_{t} }) = \bm{ w x_{t} }
\end{equation}
satisfying the maximum number of the following constraints:
\begin{equation}
\begin{array}{l}
\forall(i, j) \in O: \boldsymbol{w} \boldsymbol{x}_{\boldsymbol{i}}>\boldsymbol{w} \boldsymbol{x}_{j} \\
\forall(i, j) \in S: \boldsymbol{w} \boldsymbol{x}_{\boldsymbol{i}}=\boldsymbol{w} \boldsymbol{x}_{\boldsymbol{j}}
\end{array}
\end{equation}
where $O$ and $S$ are the ordered and similar sets respectively. It means that $O$ is composed of sample pairs $(i, j)$ with different categories, such as $i \in H$ and $j \in N$. And $S$ contains sample pairs from the same category. This setting confirms that any sample pair from the ordered set $O$ has the different emotion strength and any sample pair from the $S$ has the similar emotion strength. We believe the emotion strengths of happiness are greater than neutral.

In order to learn the weighting matrix, Parikh \textit{et. al.}~\cite{parikh2011relative} proposed to estimate the $\boldsymbol{w}$ by solving the following problem through Newton's method~\cite{chapelle2007training}:
\begin{equation}
\begin{aligned}
\operatorname{minimize} &\left(\frac{1}{2}\left\|\boldsymbol{w}_{\boldsymbol{m}}^{T}\right\|_{2}^{2}+C\left(\sum \xi_{i j}^{2}+\sum \gamma_{i j}^{2}\right)\right) \\
\text { s.t. } & \boldsymbol{w}_{\boldsymbol{m}}^{T}\left(\boldsymbol{x}_{\boldsymbol{i}}-\boldsymbol{x}_{\boldsymbol{j}}\right) \geq 1-\xi_{i j} ; \forall(i, j) \in O_{m} \\
&\left|\boldsymbol{w}_{\boldsymbol{m}}^{T}\left(\boldsymbol{x}_{\boldsymbol{i}}-\boldsymbol{x}_{\boldsymbol{j}}\right)\right| \leq \gamma_{i j} ; \forall(i, j) \in S_{m} \\
& \xi_{i j} \geq 0 ; \gamma_{i j} \geq 0
\end{aligned}
\end{equation}
where $C$ is utilized to control the trade-off between the margin and the size of the slack variables $\xi_{i j}$ and $\gamma_{i j}$.

\begin{figure}[ht]
        \centering
        \includegraphics[width=1.0\linewidth]{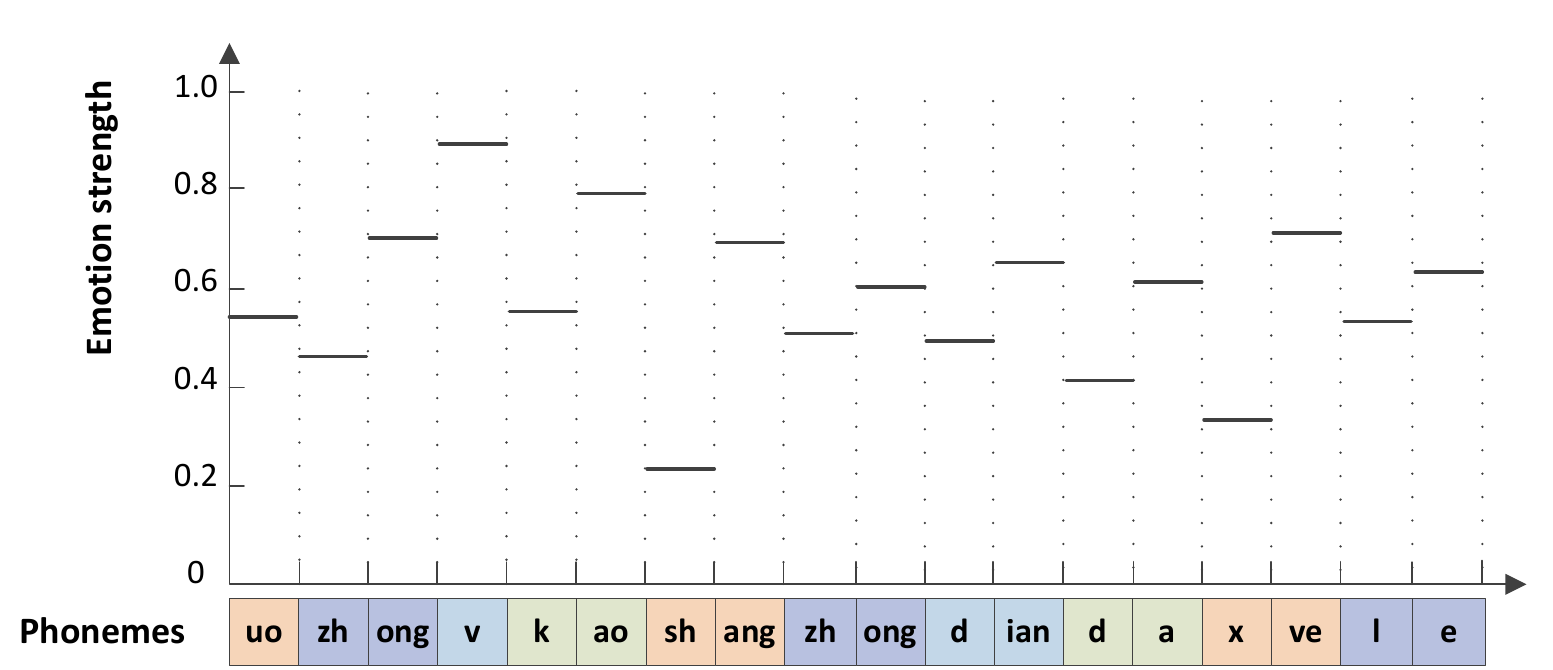}
        \caption{Phoneme-level emotion strength representation of a ``happy'' utterance}
         \vspace{-10pt}
        \label{fig:strength}
\end{figure}

To extract the phoneme-level emotion strength for fine-grained synthesis, we firstly train an HMM-based alignment model to obtain the phoneme boundaries of each utterance. With the learned ranking weights $\boldsymbol{w}$ and $N$ speech fragments $\{x_{i}^1,x_{i}^2,...,x_{i}^N\}$ according to the boundaries, we could obtain the phoneme-level emotion strength from Eq.~\eqref{eq:1}, which are treated as the local emotion descriptors. We finally normalize the emotion strength into $[0,1]$ in each emotion category for easy-to-use in emotion control. Fig.~\ref{fig:strength} shows an utterance with phoneme-level emotion strength. We can see that different phonemes have different local emotion strengths, although they all have the same utterance-level emotion render -- happy.

\subsection{Fine-grained emotion transfer and control}
\label{subsec:transfer}
In the emotion transfer or control scenario, the generated speech is forced to perform the emotion expression of the reference audio or manual labels. With the above phoneme-level emotion strength representation and global emotion render, we firstly build an attention-based seq2seq model for emotional speech synthesis, as shown in the training part of Fig.~\ref{fig:sys}. In this model, the emotion category embedding is treated as the global emotion render, and the phoneme-level strength represents the local emotion descriptors.

Considering emotion transfer, we can utilize a force-alignment model to obtain the phoneme boundaries of the target reference audio. Then the fragments of the reference audio can be adopted to compute the phoneme-level emotion strength through the learned ranking function. Since the phoneme number of reference audio is different from the input text during inference, we need to conduct linear interpolation or down-sampling to obtain the phoneme-level strength sequence whose length is the same as the text phoneme sequence. In details, assuming there are $M$ phonemes in the reference audio, we can simply construct a curve through the interpolation of $M$ local emotion descriptors. When the input text contains $N$ phonemes, we evenly split the above curve and treat the boundaries as the target local emotion descriptors.

As for emotion control, the model needs to perform according to manual instructions. Given the input phoneme sequence and the emotion category during inference, we can directly design the local emotion descriptors to satisfy our needs. That is to say, we can assign any value in [0, 1] for each phoneme of the input text to control the generated audio into any trend of emotion expressions as needed. In this way, the proposed model can use the manual designed emotion labels (strength) to obtain fine-grained and flexible control.

\subsection{Fine-grained emotion prediction}
As discussed above, emotion transfer or control needs reference audio or manual labels to decide the emotional expressions of synthesized speech. But in practice, it is hard to find suitable reference audio or manually-designed emotion labels at the phoneme level. Thus in the proposed fine-grained emotion prediction module, we directly predict phoneme-level local descriptors from phoneme sequences. So the text encoder needs to provide content information for the acoustic decoder and predicted emotion strength information for each phoneme at the same time. As a result, the proposed model can produce natural emotional speech without any reference audio or manual label. As mentioned in section~\ref{subsec:transfer}, we can also conduct emotion transfer and control with this unified model.

In detail, we feed the encoder output to a strength predictor, which has two fully-connected layers followed by the ReLU activation, to predict emotion strength. We minimize the differences between the predicted strength and the ground-truth strength extracted from the relative attributes ranking function. So the final objective of acoustic model is:
\begin{equation}\label{eq:2}
Loss = Loss_{mel} + \alpha Loss_{strength}
\end{equation}
where $Loss_{mel}$ means the conventional L1 loss for acoustic modeling, and $Loss_{strength}$ is the L1 loss for emotion strength. $\alpha$ is a tunable weight during training.

During inference, our model will predict the phoneme-level emotion strength directly from text without any reference or label. And the predicted strength will decide the emotional expressions in the generated speech, given the emotion category as a global render. Since the phoneme-level emotion prediction module is relatively independent, we can also use phoneme-level emotion strength from reference audio or manual labels in the same model, which means that the proposed unified model is flexible for emotion transfer, control and prediction.

\section{Experiments}
\subsection{Basic setups}
In our experiments, we use an internal high-quality emotional speech corpus, which contains about 14-hours of speech from a professional Chinese female speaker. The corpus consists of about 6000 sentences of neutral speech and six categories of emotional speech, including happy, angry, disgust, fear, surprise and sad, where each emotion category has about 600 sentences.

For text representation, we analyze the phone, tone and prosody boundary information through our text analysis module. We extract 80-band mel-scale spectrogram from speech as acoustic features. For both objective and subjective evaluation, we reserve 30 sentences of each emotional category to evaluate the performance of emotional style transfer and control. To reconstruct waveforms from mel-spectrogram, we build a multi-band WaveRNN~\cite{yu2019durian} trained by ground-truth mel-spectrogram for fair comparisons. There are 20 native Chinese speakers taking part in the subjective evaluation. And for the objective evaluation, dynamic time warping (DTW) is adopted to align predicted features and target features.

\subsection{Model details}

For the fine-grained local emotion descriptors, we firstly extract 384-dimensional emotion-related features from speech using the openSMILE tool~\cite{eyben2010opensmile} and learn the ranking function using the MATLAB codes provided by Parikh \textit{et al.}~\cite{parikh2011relative}. We train an HMM-based aligner using the same corpus to obtain the phoneme boundary of each utterance. Finally, we utilize the phoneme-level speech fragments to compute the emotion strength as local emotion descriptors through the learned ranking function. The extracted phoneme-level descriptors are finally normalized into [0,1].

\begin{table}[!thb]
  \caption{MCD of different models for parallel transfer}
  \vspace{-10pt}
  \label{tab:paral}
  \centering
  \begin{tabular}{c c c c}
    \toprule
    \multicolumn{1}{c}{\textbf{Method}} &\multicolumn{1}{c}{\textbf{MCD (dB)}} \\
    \midrule
    GST & 4.89\\
    UET & 5.16\\
    proposed-FET & 4.91\\
    \bottomrule
  \end{tabular}
\end{table}

\begin{figure}[!thb]
        \centering
        \includegraphics[width=1.0\linewidth]{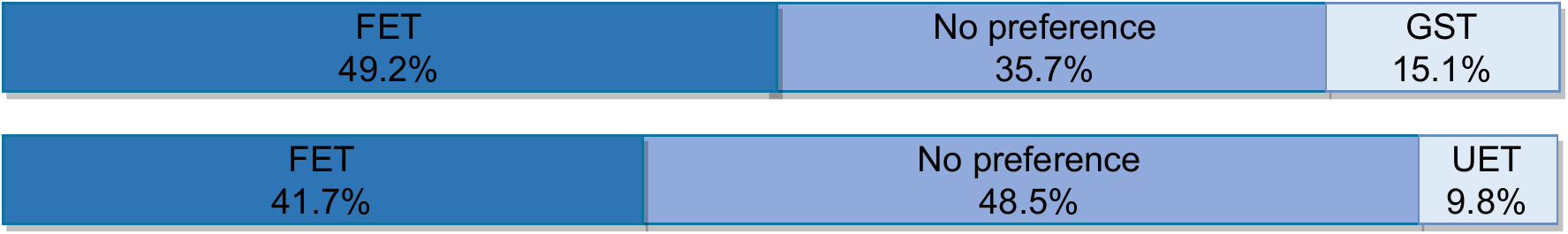}
        \caption{A/B preference test result for parallel transfer}
        \label{fig:paral-abx}
\end{figure}

\begin{figure}[!thb]
        \centering
        \includegraphics[width=1.0\linewidth]{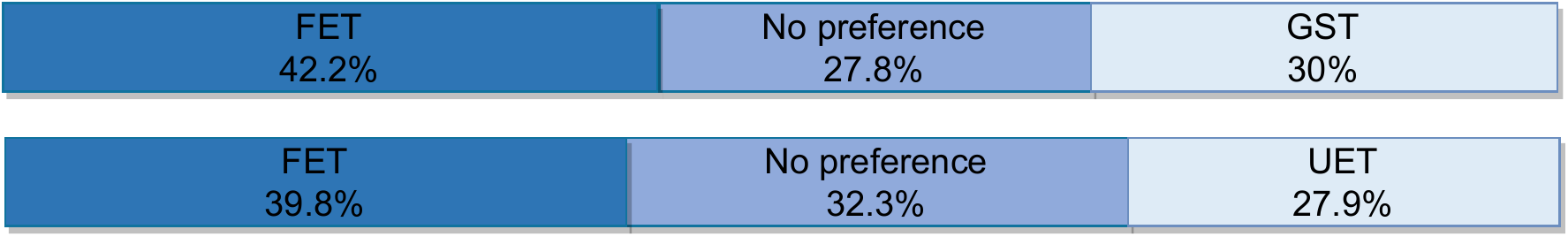}
        \caption{A/B preference test for non-parallel transfer}
         \vspace{-10pt}
        \label{fig:non-paral-abx}
\end{figure}

In the emotional speech synthesis model,  we adopt three feed-forward layers as pre-net followed by a CBHG module as text encoder~\cite{wang2017tacotron}. For the fine-grained emotion control, we concatenate the emotional category embedding with the encoder outputs as global emotion render, while the phoneme-level emotion descriptors are added to the encoder outputs through a linear projection layer with 512 units. As for the emotion prediction, the encoder outputs are fed into the prediction network to predict the phoneme-level emotion strengths. The auto-regressive decoder also contains three feed-forward layers as pre-net and a 2048 units unidirectional LSTM layer. We adopt the robust GMM attention~\cite{battenberg2020location} to connect the encoder and the decoder. Finally, the CBHG-based post-net is used to produce mel-spectrogram and waveform is generated through multiband WaveRNN.

\subsection{Experimental results}
Through our proposed approach, we can conduct fine-grained emotion transfer, control and prediction in a unified model. Hence we will evaluate the performance of the proposed model in the three aspects.

\subsubsection{Fine-grained emotion transfer}
We first evaluate the ability of emotion transfer for the proposed fine-grained emotion transfer (FET). We train the GST model~\cite{wang2018style} and the utterance-level emotion transfer model~\cite{zhu2019controlling} (UET) as baseline systems for comparison. We evaluate the performance of both parallel and non-parallel transfer. For the parallel transfer, the reference audio has the same text content as the target text to be synthesized. In this scenario, the mel-spectrogram of reference audio is fed into the reference encoder of the GST model, and we extract utterance-level and phoneme-level emotion strength of reference audio for the UET model and proposed FET model respectively. As for the non-parallel transfer, where the target text is not necessarily the same as that of the reference audio, the reference audios are randomly selected from the test set to conduct emotion transfer.

Table~\ref{tab:paral} shows the mel-cepstral distortion (MCD) of different models for parallel transfer. The results indicate that the GST model and the proposed FET model have apparently lower MCD values as compared with the utterance-level emotional strength model.  The proposed FET model has a close MCD value with the GST model. Since the goal of emotion transfer is to imitate the emotion of the reference speech, we also conduct A/B preference test to let listeners choose which one is more similar to the reference in emotion expressions, as shown in Fig.~\ref{fig:paral-abx}. Comparing the proposed FET with the GST model, we find that with much more preferred, the FET model can imitate the local emotional expressions better than the GST model which only can learn an ``averaged'' emotional embedding from reference. Similarly, the performance of phoneme-level imitation (FET) is much better than the utterance-level transfer (UET).

\begin{figure}[!htb]
\vspace{-10pt}
\centering
\subfigure[Disgust]{
    \begin{minipage}[t]{0.22\textwidth}
    \includegraphics[width=4.5cm]{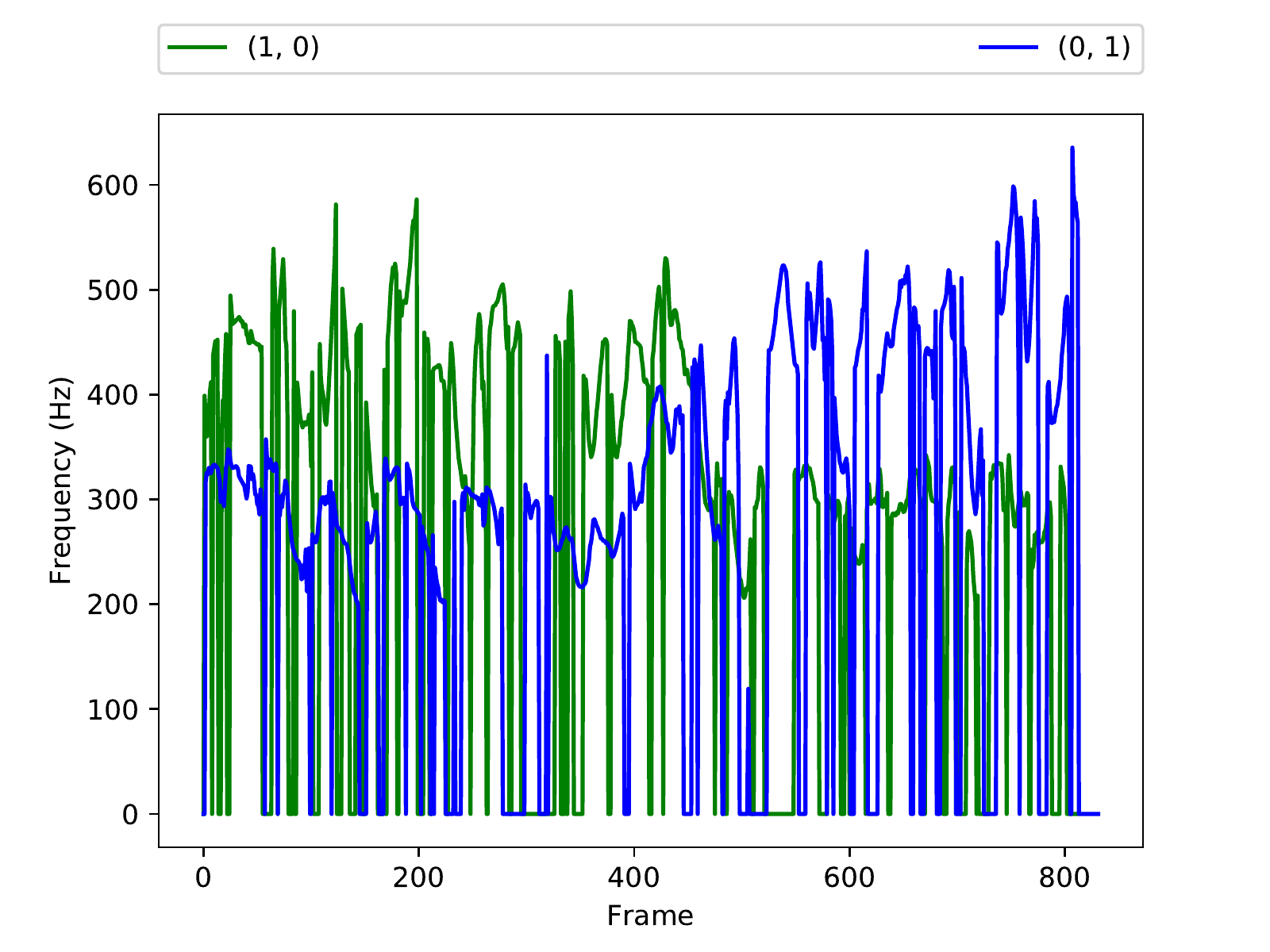}
    \end{minipage}
}
\subfigure[Fear]{
    \begin{minipage}[t]{0.22\textwidth}
    \includegraphics[width=4.5cm]{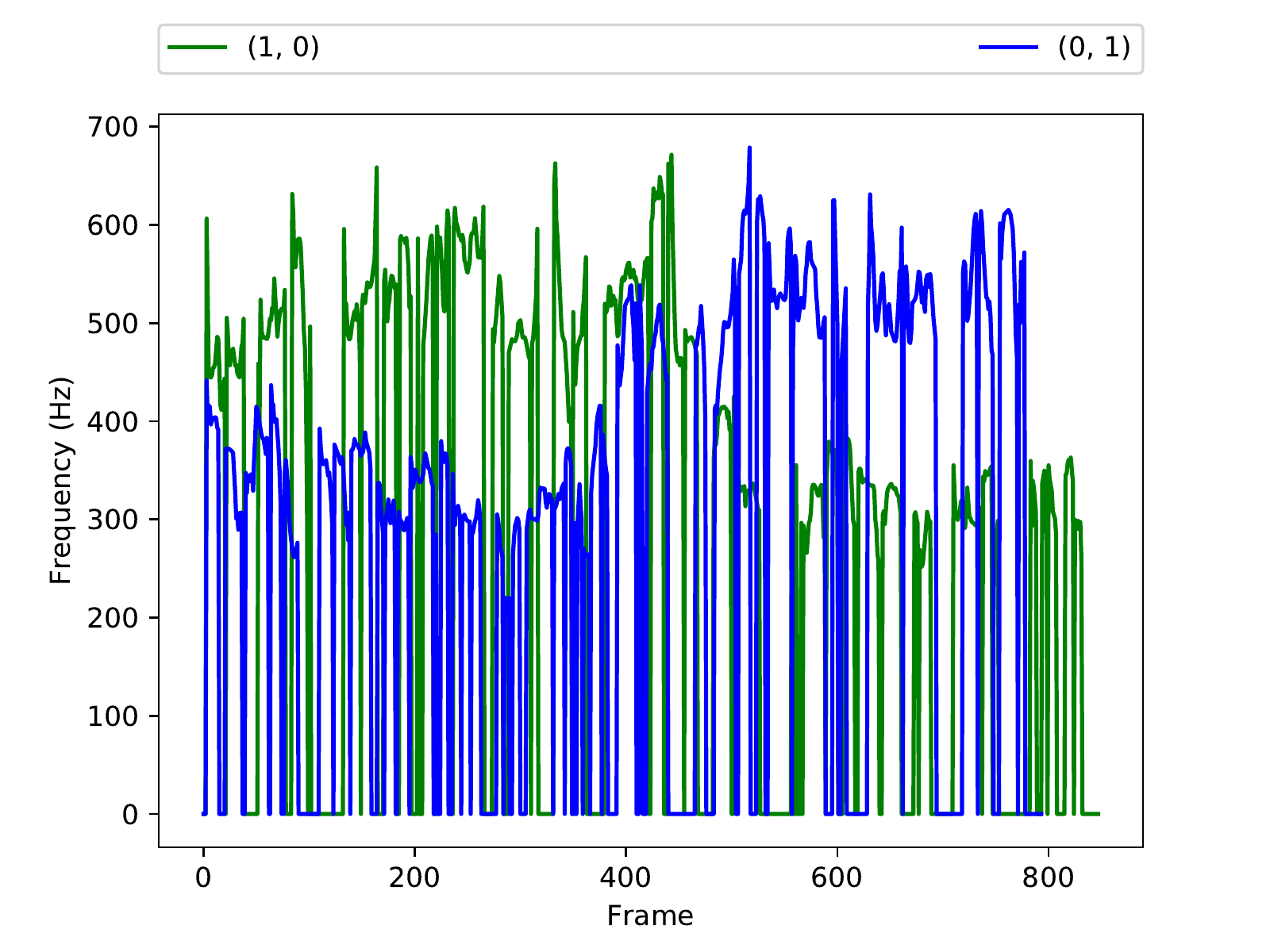}
    \end{minipage}
}
\vspace{-10pt}
\caption{F0 curves of synthetic samples of different global render and local descriptors. (0,1) means that the local descriptors of the first half phonemes is 0, and the last half is 1.}
\label{fig:f0}
\end{figure}

\begin{figure}[!htb]
\centering
\subfigure[Gradually decreased strength]{
    \begin{minipage}[t]{0.3\textwidth}
    \includegraphics[width=5.5cm]{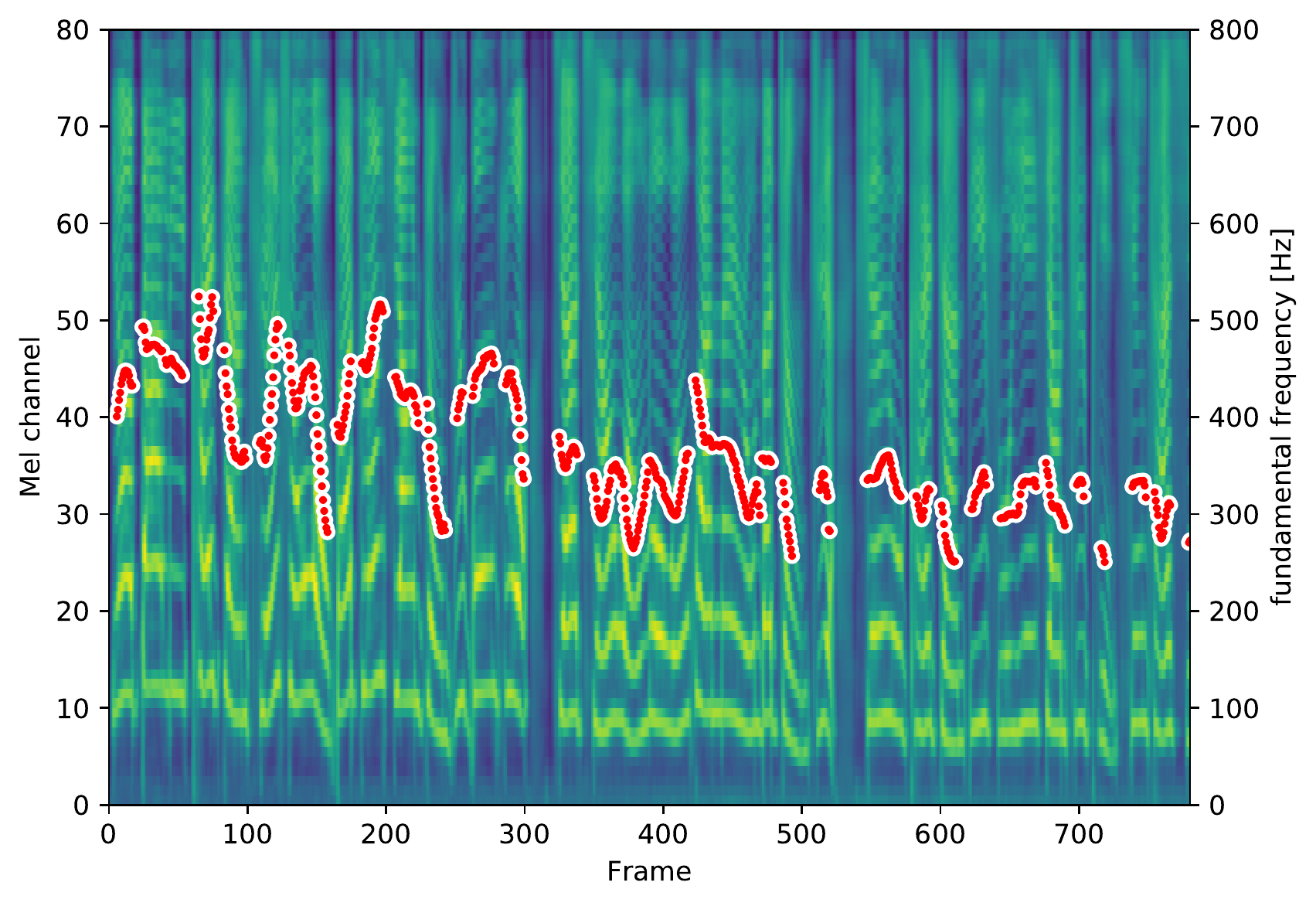}
    \end{minipage}
}
\subfigure[Gradually increased strength]{
    \begin{minipage}[t]{0.3\textwidth}
    \includegraphics[width=5.5cm]{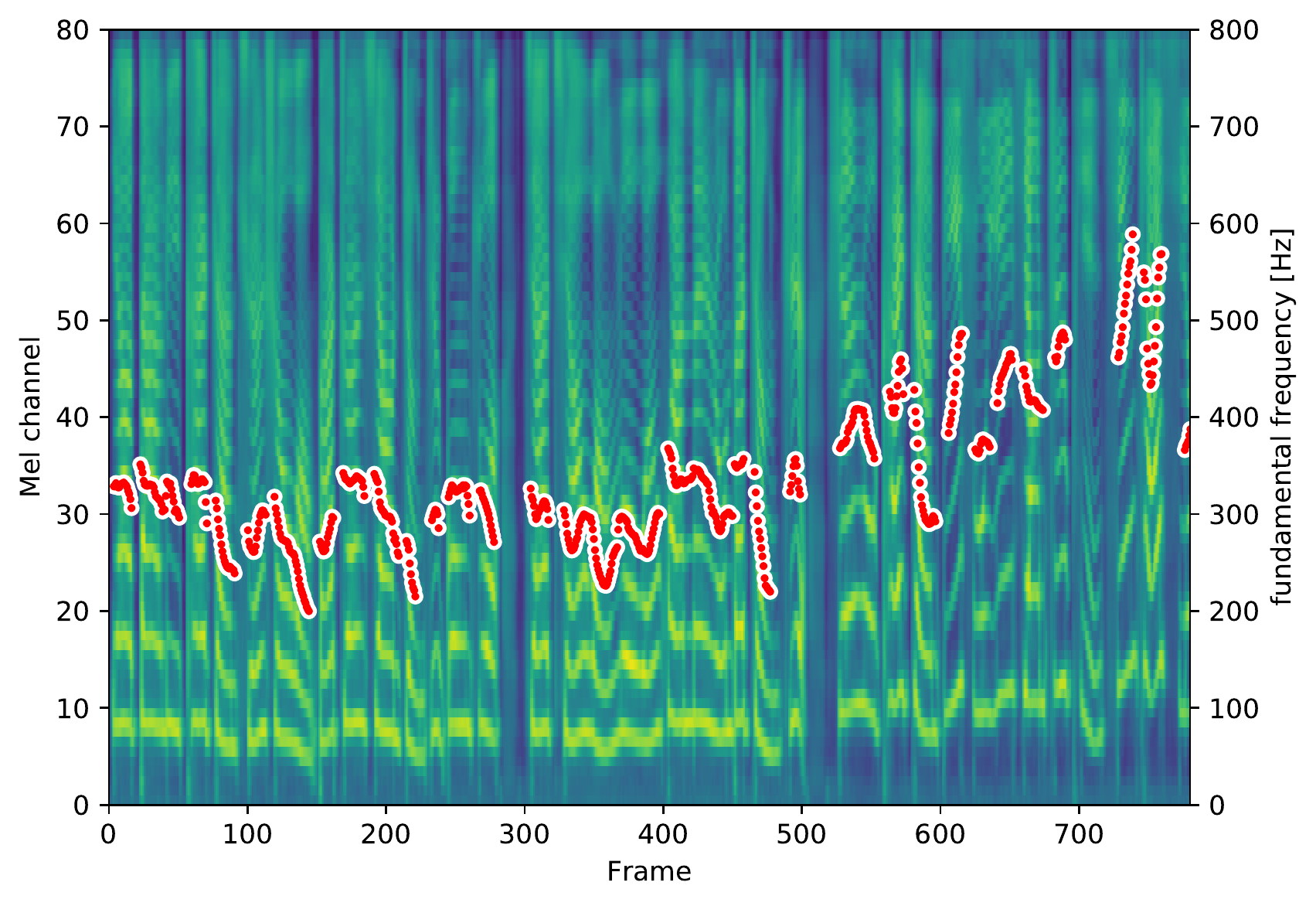}
    \end{minipage}
}
\vspace{-10pt}
\caption{Mel spectrograms and F0 of synthetic samples with gradually changed emotion strength.}
\label{fig:mel}
\end{figure}

\begin{figure}[!htb]
        \centering
        \includegraphics[width=1.0\linewidth]{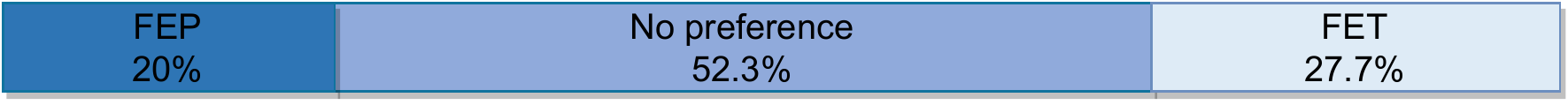}
        \caption{A/B preference for FEP and FET}
        \label{fig:pred-abx}
\vspace{-2pt}
\end{figure}

\label{sec:ref}
\begin{figure*}[th]
\vspace{-10pt}
\centering
\subfigure[Recording]{
    \begin{minipage}[b]{0.32\textwidth}
    \includegraphics[width=5.5cm]{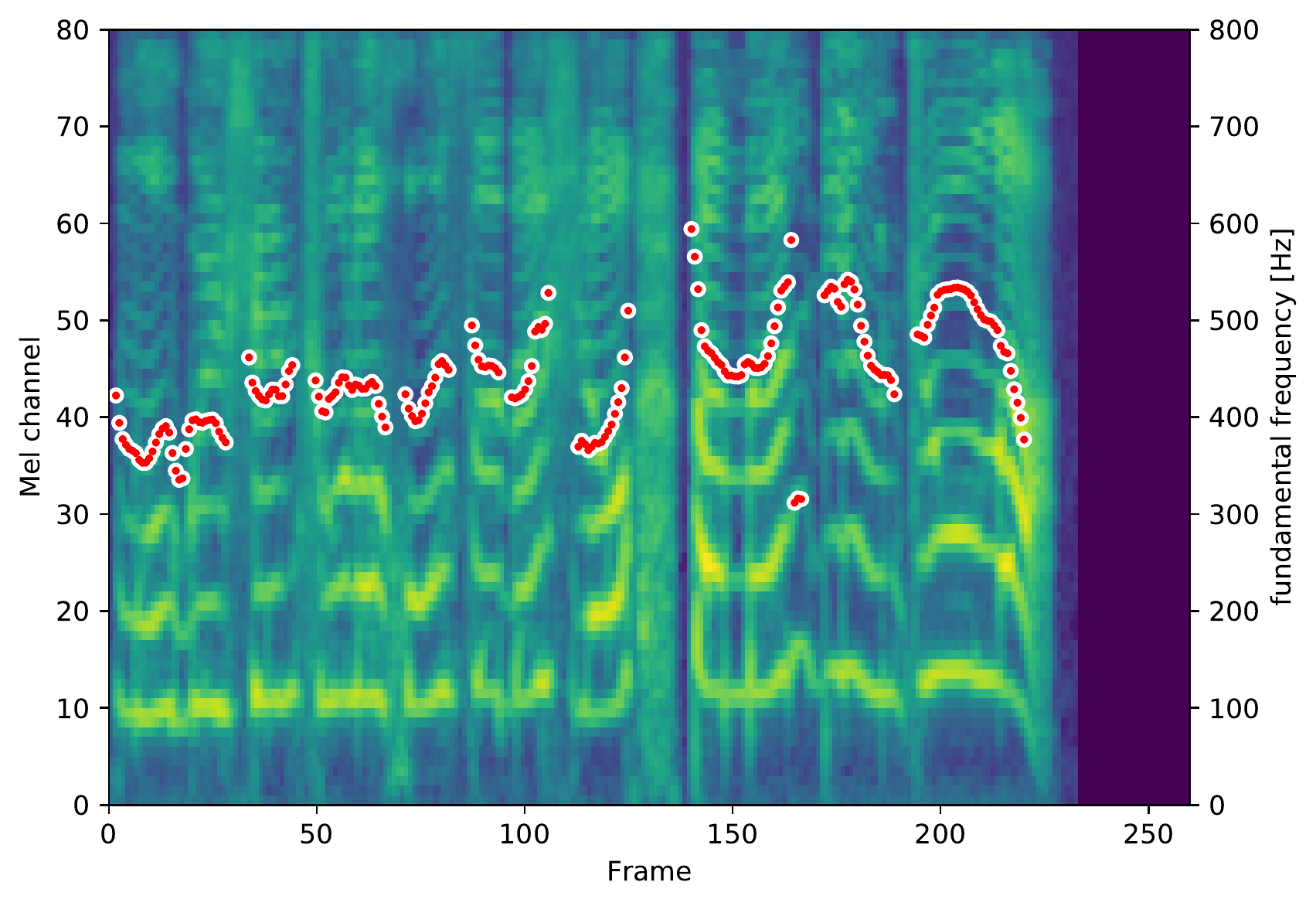}
    \end{minipage}
}
\subfigure[Parallel transfer]{
    \begin{minipage}[b]{0.32\textwidth}
    \includegraphics[width=5.5cm]{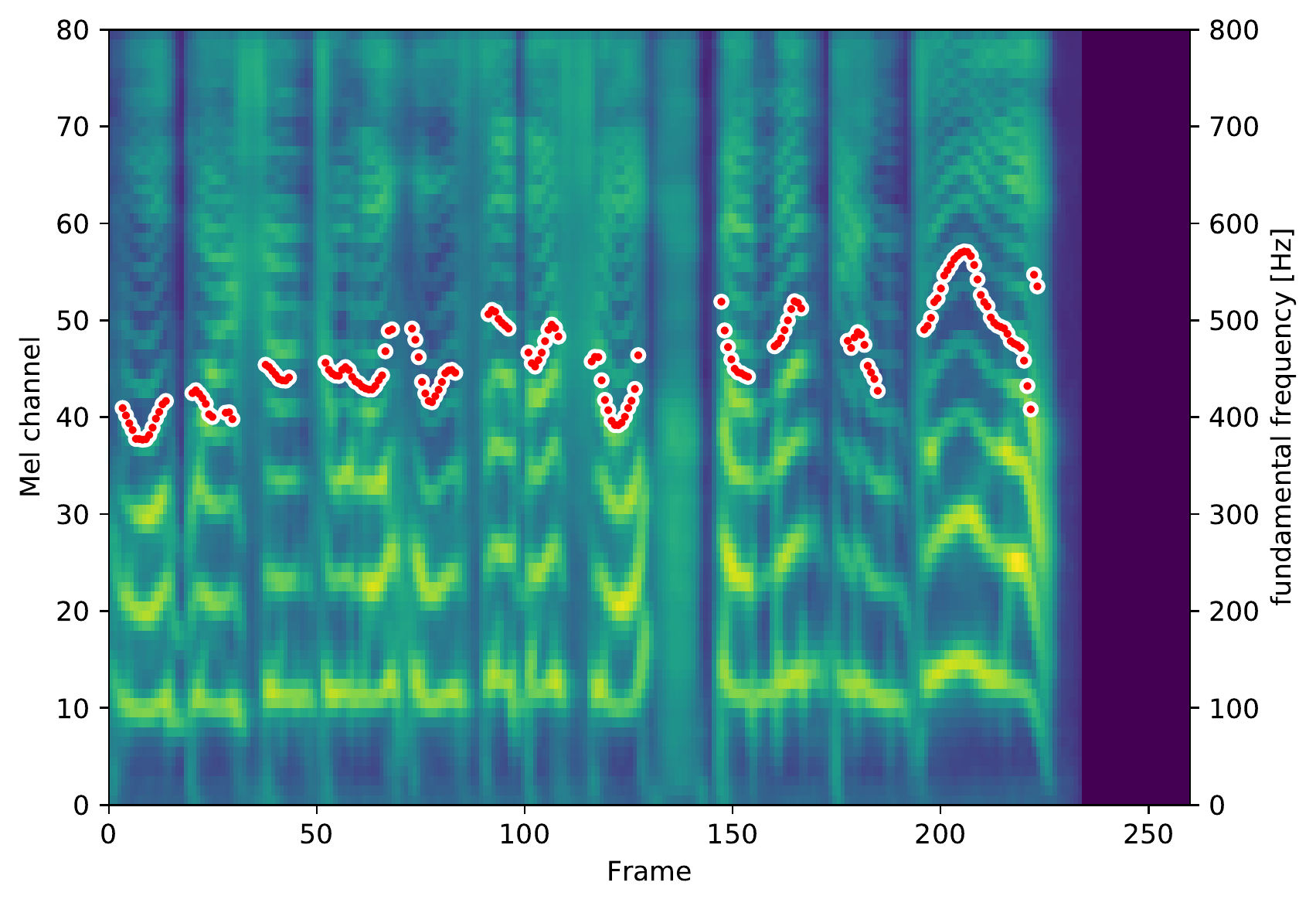}
    \end{minipage}
}
\subfigure[Prediction]{
    \begin{minipage}[b]{0.32\textwidth}
    \includegraphics[width=5.5cm]{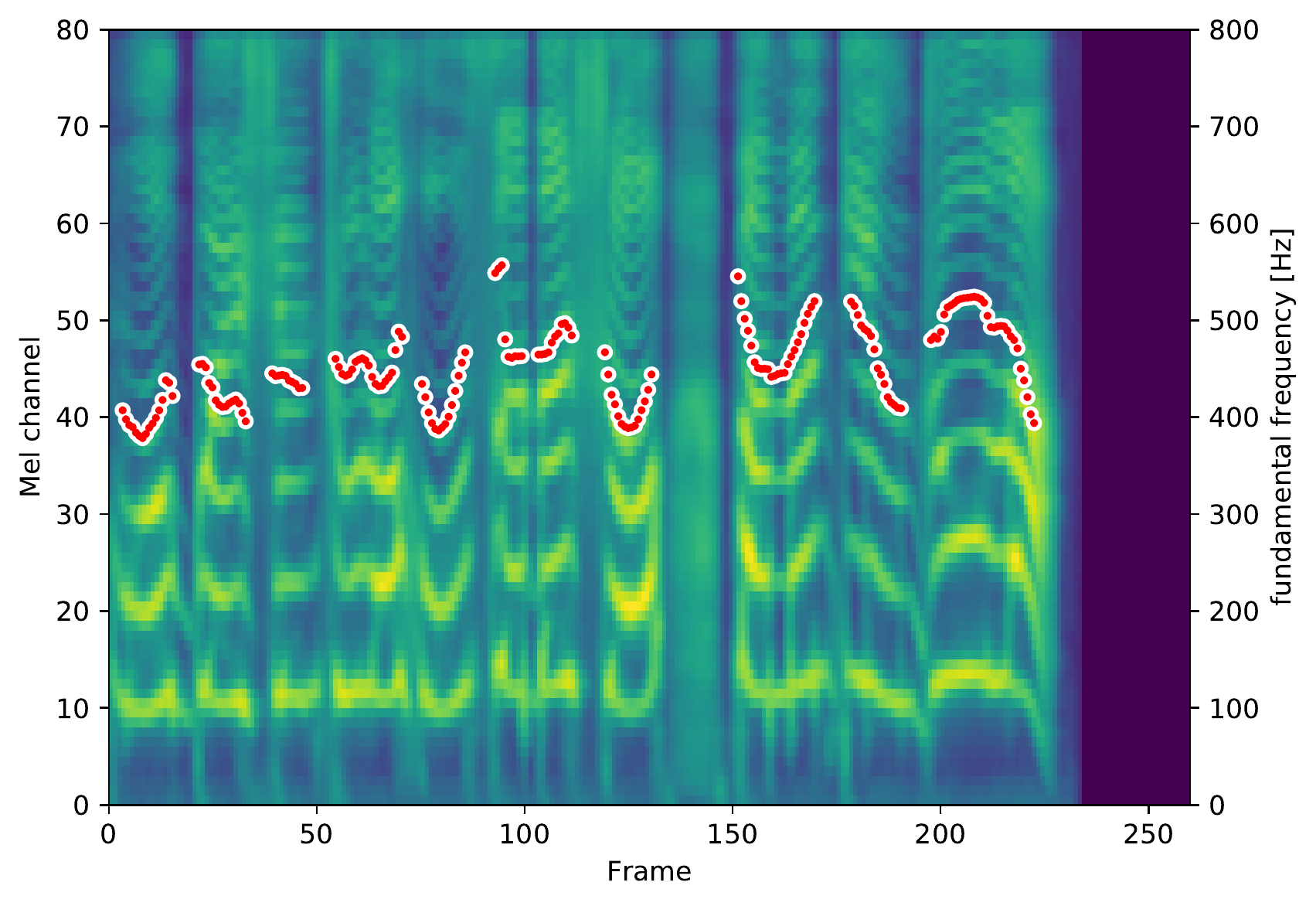}
    \end{minipage}
}
\vspace{-10pt}
\caption{Mel spectrograms and F0 from recording, parallel emotion transfer and emotion prediction model.}
\vspace{-10pt}
\label{fig:pre.vs.parall}
\end{figure*}

We also evaluate the performance of non-parallel transfer for the above models. The results of A/B preference test are shown in Fig.~\ref{fig:non-paral-abx}. As for the non-parallel transfer, we find the proposed FET model also outperforms the GST and UET models.

\subsubsection{Fine-grained emotion control}
Another important aspect of emotional speech synthesis is how to flexibly control the emotional expressions of synthesized speech. For the GST model, we can choose or weigh different tokens to roughly control the speech generation. And the UET model can also control the emotion render with different scales, but only at the whole-utterance level.  By contrast, the proposed FET model has the ability of fine controls for both global render and local descriptor. To show such ability, we generate utterances with different global render (emotion category) and local descriptors (phoneme-level emotion strength). Fig.~\ref{fig:f0} shows two examples of F0 trajectory, with a ``disgust'' render and a ``fear'' render. Note that F0 is directly related to emotion strength. In each emotional render, we split the phoneme sequence of the text into two parts, where the emotion strengths of the phonemes in each part are set as either 0 or 1. The F0 trajectories in Fig.~\ref{fig:f0} show that the utterance part whose emotion strengths are 1 have much higher F0 in the generated speech as compared to the part with strengths of 0. Besides, we also set the strength of the input phoneme sequence gradually decrease from 1 to 0 and gradually increase from 0 to 1 for obtaining gradual strength change. Fig.~\ref{fig:mel} shows that the F0 trajectory changes gradually as our control. Subjective listening on these samples also indicates that fine-grained emotion strength change can be easily detected. We suggest the readers listen to our online demos\footnote{Samples can be found at \url{https://leiyi420.github.io/pho_ra_strength/}}.

\subsubsection{Fine-grained emotion prediction}

For the above emotion transfer and control, we still need auxiliary information from reference audio or manual setup to control the emotional expressions. We finally evaluate the proposed fine-grained emotion prediction model (FEP), which directly predicts phoneme-level local emotion strengths from text. Given the emotion category, we generate fine-grained emotional speech with the predicted local emotion strength and calculate the MCD with the target emotional speech, as shown in Table~\ref{tab:pre}. From the results, we can find that the generated speech with the ground-truth phoneme strength (FET) has slightly lower MCD than the one using predicted strength (FEP), but the difference is not big. We also conduct A/B test on the FET and FEP systems, as shown in Fig.~\ref{fig:pred-abx}. The subjective result indicates that there is no significant difference between generated speech with predicted strength and ground-truth strength, which shows the ability of the proposed FEP model to accurately predict the local emotional descriptors.

  \vspace{-5pt}
\begin{table}[!thb]
  \caption{MCD of emotion prediction and parallel transfer}
  \vspace{-10pt}
  \label{tab:pre}
  \centering
  \begin{tabular}{c c c c}
    \toprule
    \multicolumn{1}{c}{\textbf{Method}} &\multicolumn{1}{c}{\textbf{MCD (dB)}} \\
    \midrule
    proposed-FET & 4.91\\
    proposed-FEP & 5.03\\
    \bottomrule
  \end{tabular}
\vspace{-10pt}
\end{table}

In order to intuitively see the difference between the predicted emotion expression and parallel emotion transfer, we also analyze the mel-spectrogram and F0 of the generated speech from both models. Fig.~\ref{fig:pre.vs.parall} shows the generated examples from the parallel transfer method, prediction method and real recording. We can find that the synthesized speech using predicted phoneme-level strength can reconstruct satisfied emotion expressions like parallel emotion transfer, even though there are some differences in a few units. The results prove that the proposed method can predict similar emotion expression to both parallel transfer and recordings directly from arbitrary text, which does not need reference audio or manual interventions.

\section{Conclusions}

This paper proposed a unified model to conduct emotion transfer, control and prediction for fine-grained emotional speech synthesis. With a sentence-level global render of the emotion category and the phoneme-level local descriptors of the emotion strength learned from a ranking function, the proposed model can transfer details of emotion intensities from reference audio and synthesize speech from manual labels to control emotion expressions. In addition, it can also predict phoneme-level emotional expressions directly from texts. Experimental results show that the proposed method can transfer, control and predict the fine-grained emotion expression in a unified model, which outperforms the baseline systems with coarse emotional expressions and also improves the flexibility of emotional speech synthesis model.


\bibliographystyle{IEEEbib}
\bibliography{strings,refs}

\begin{thebibliography}{10}

\bibitem{ze2013statistical}
Heiga Zen, Andrew Senior, and Mike Schuster,
\newblock ``Statistical parametric speech synthesis using deep neural
  networks,''
\newblock in {\em Proc. ICASSP}. IEEE, 2013, pp. 7962--7966.

\bibitem{qian2014training}
Yao Qian, Yuchen Fan, Wenping Hu, and Frank~K Soong,
\newblock ``On the training aspects of deep neural network (dnn) for parametric
  tts synthesis,''
\newblock in {\em Proc. ICASSP}. IEEE, 2014, pp. 3829--3833.

\bibitem{ling2013modeling}
Zhen-Hua Ling, Li~Deng, and Dong Yu,
\newblock ``Modeling spectral envelopes using restricted boltzmann machines and
  deep belief networks for statistical parametric speech synthesis,''
\newblock {\em IEEE transactions on audio, speech, and language processing},
  vol. 21, no. 10, pp. 2129--2139, 2013.

\bibitem{wang2017tacotron}
Yuxuan Wang, RJ~Skerry-Ryan, Daisy Stanton, Yonghui Wu, Ron~J Weiss, Navdeep
  Jaitly, Zongheng Yang, et~al.,
\newblock ``{Tacotron: Towards end-to-end speech synthesis},''
\newblock in {\em Proc. INTERSPEECH}, 2017, pp. 4006--4010.

\bibitem{shen2018natural}
Jonathan Shen, Ruoming Pang, Ron~J Weiss, Mike Schuster, Navdeep Jaitly,
  Zongheng Yang, Zhifeng Chen, Yu~Zhang, Yuxuan Wang, Rj~Skerrv-Ryan, et~al.,
\newblock ``Natural tts synthesis by conditioning wavenet on mel spectrogram
  predictions,''
\newblock in {\em Proc. ICASSP}. IEEE, 2018, pp. 4779--4783.

\bibitem{li2019neural}
Naihan Li, Shujie Liu, Yanqing Liu, Sheng Zhao, and Ming Liu,
\newblock ``Neural speech synthesis with transformer network,''
\newblock in {\em Proc. AAAI}, 2019, vol.~33, pp. 6706--6713.

\bibitem{yang2020localness}
Shan Yang, Heng Lu, Shiyin Kang, Liumeng Xue, Jinba Xiao, Dan Su, Lei Xie, and
  Dong Yu,
\newblock ``On the localness modeling for the self-attention based end-to-end
  speech synthesis,''
\newblock {\em Neural Networks}, vol. 125, pp. 121--130, 2020.

\bibitem{skerry2018towards}
RJ~Skerry-Ryan, Eric Battenberg, Ying Xiao, Yuxuan Wang, Daisy Stanton, Joel
  Shor, Ron~J Weiss, Rob Clark, and Rif~A Saurous,
\newblock ``Towards end-to-end prosody transfer for expressive speech synthesis
  with tacotron,''
\newblock {\em arXiv preprint arXiv:1803.09047}, 2018.

\bibitem{wang2018style}
Yuxuan Wang, Daisy Stanton, Yu~Zhang, RJ~Skerry-Ryan, Eric Battenberg, Joel
  Shor, Ying Xiao, Fei Ren, Ye~Jia, and Rif~A Saurous,
\newblock ``Style tokens: Unsupervised style modeling, control and transfer in
  end-to-end speech synthesis,''
\newblock {\em arXiv preprint arXiv:1803.09017}, 2018.

\bibitem{stanton2018predicting}
Daisy Stanton, Yuxuan Wang, and RJ~Skerry-Ryan,
\newblock ``Predicting expressive speaking style from text in end-to-end speech
  synthesis,''
\newblock in {\em Proc. SLT}. IEEE, 2018, pp. 595--602.

\bibitem{zhang2019learning}
Ya-Jie Zhang, Shifeng Pan, Lei He, and Zhen-Hua Ling,
\newblock ``Learning latent representations for style control and transfer in
  end-to-end speech synthesis,''
\newblock in {\em Proc. ICASSP}. IEEE, 2019, pp. 6945--6949.

\bibitem{bian2019multi}
Yanyao Bian, Changbin Chen, Yongguo Kang, and Zhenglin Pan,
\newblock ``Multi-reference tacotron by intercross training for style
  disentangling, transfer and control in speech synthesis,''
\newblock {\em arXiv preprint arXiv:1904.02373}, 2019.

\bibitem{henter2018deep}
Gustav~Eje Henter, Jaime Lorenzo-Trueba, Xin Wang, and Junichi Yamagishi,
\newblock ``Deep encoder-decoder models for unsupervised learning of
  controllable speech synthesis,''
\newblock {\em arXiv preprint arXiv:1807.11470}, 2018.

\bibitem{um2020emotional}
Se-Yun Um, Sangshin Oh, Kyungguen Byun, Inseon Jang, ChungHyun Ahn, and
  Hong-Goo Kang,
\newblock ``Emotional speech synthesis with rich and granularized control,''
\newblock in {\em Proc.ICASSP}. IEEE, 2020, pp. 7254--7258.

\bibitem{lee2019robust}
Younggun Lee and Taesu Kim,
\newblock ``Robust and fine-grained prosody control of end-to-end speech
  synthesis,''
\newblock in {\em Proc. ICASSP}. IEEE, 2019, pp. 5911--5915.

\bibitem{ferrari2008learning}
Vittorio Ferrari and Andrew Zisserman,
\newblock ``Learning visual attributes,''
\newblock in {\em Proc. NeurIPS}, 2008, pp. 433--440.

\bibitem{parikh2011relative}
Devi Parikh and Kristen Grauman,
\newblock ``Relative attributes,''
\newblock in {\em Proc. ICCV}. IEEE, 2011, pp. 503--510.

\bibitem{zhu2019controlling}
Xiaolian Zhu, Shan Yang, Geng Yang, and Lei Xie,
\newblock ``Controlling emotion strength with relative attribute for end-to-end
  speech synthesis,''
\newblock in {\em Proc. ASRU}. IEEE, 2019, pp. 192--199.

\bibitem{lorenzo2018investigating}
Jaime Lorenzo-Trueba, Gustav~Eje Henter, Shinji Takaki, Junichi Yamagishi,
  Yosuke Morino, and Yuta Ochiai,
\newblock ``Investigating different representations for modeling and
  controlling multiple emotions in dnn-based speech synthesis,''
\newblock {\em Speech Communication}, vol. 99, pp. 135--143, 2018.

\bibitem{lee2017emotional}
Younggun Lee, Azam Rabiee, and Soo-Young Lee,
\newblock ``Emotional end-to-end neural speech synthesizer,''
\newblock {\em arXiv preprint arXiv:1711.05447}, 2017.

\bibitem{chapelle2007training}
Olivier Chapelle,
\newblock ``Training a support vector machine in the primal,''
\newblock {\em Neural computation}, vol. 19, no. 5, pp. 1155--1178, 2007.

\bibitem{yu2019durian}
Chengzhu Yu, Heng Lu, Na~Hu, Meng Yu, Chao Weng, Kun Xu, Peng Liu, Deyi Tuo,
  Shiyin Kang, Guangzhi Lei, et~al.,
\newblock ``Durian: Duration informed attention network for multimodal
  synthesis,''
\newblock {\em arXiv preprint arXiv:1909.01700}, 2019.

\bibitem{eyben2010opensmile}
Florian Eyben, Martin W{\"o}llmer, and Bj{\"o}rn Schuller,
\newblock ``Opensmile: the munich versatile and fast open-source audio feature
  extractor,''
\newblock in {\em Proc. ACMMM}, 2010, pp. 1459--1462.

\bibitem{battenberg2020location}
Eric Battenberg, RJ~Skerry-Ryan, Soroosh Mariooryad, Daisy Stanton, David Kao,
  Matt Shannon, and Tom Bagby,
\newblock ``Location-relative attention mechanisms for robust long-form speech
  synthesis,''
\newblock in {\em Proc. ICASSP}. IEEE, 2020, pp. 6194--6198.

\end{thebibliography}

\end{document}